\documentclass[amsmath,eqsecnum,preprint,pre,aps,showpacs]{revtex4}
\usepackage{amsmath}
\usepackage{graphicx}

\begin{document}

\title{A New Theory of Dynamic Arrest in Colloidal Mixtures}
\author{R. Ju\'arez-Maldonado and M. Medina-Noyola}

\address{Instituto de F\'{\i}sica {\sl ``Manuel Sandoval
Vallarta"}, Universidad Aut\'{o}noma de San Luis Potos\'{\i},
\'{A}lvaro Obreg\'{o}n 64, 78000 San Luis Potos\'{\i}, SLP,
M\'{e}xico}
\date{\today}

\begin{abstract}

We present a new first-principles theory of dynamic arrest in
colloidal mixtures based on the multi-component self-consistent
generalized Langevin equation (SCGLE) theory of  colloid dynamics
[Phys. Rev. E {\bf 72}, 031107 (2005); ibid {\bf 76}, 039902
(2007)]. We illustrate its application with the description of
dynamic arrest in two simple model colloidal mixtures, namely, the
hard-sphere and the repulsive Yukawa binary mixtures. Our results
include the observation of the two patterns of dynamic arrest, one
in which both species become simultaneously arrested, and the other
involving the sequential arrest of the two species. The latter case
gives rise to mixed states in which one species is arrested while
the other species remains mobile. We also derive the (``fixed
point") equations for the non-ergodic parameters of the system,
which takes the surprisingly simple form of a system of coupled
equations for the localization length of the particles of each
species. The solution of this system of equations indicates
unambiguously which species is arrested (finite localization length)
and which species remains ergodic (infinite localization length). As
a result, we are able to draw the entire ergodic--non-ergodic phase
diagram of the binary hard-sphere mixture.

\end{abstract}


\maketitle

\section{Introduction}

The fundamental understanding of dynamically arrested states is a
major challenge of contemporary statistical physics and materials
science \cite{angell, debenedetti, goetze1}. Model experimental
colloidal suspensions, whose dynamics has been the subject of study
on its own right \cite{1,6,2}, have played an essential role in the
study of dynamic arrest phenomena, providing experimental
realizations in finely-controlled systems and conditions
\cite{vanmegen1, bartsch, beck, pham, sciortinotartaglia}. Thus, it
is an experimental fact that by changing a macroscopic control
parameter such as the volume fraction, or by tuning the effective
inter-particle interactions, one may drive the system from the
region of equilibrium (ergodic) states to the region of dynamically
arrested (non-ergodic) states. First-principles models and theories
that predict and describe these transitions thus constitute an
essential aspect of the fundamental understanding of these
phenomena. The mode coupling theory (MCT) of the ideal glass
transition \cite{goetze1, goetze2} is perhaps the most comprehensive
theory of this sort, some of whose predictions have found beautiful
experimental confirmation \cite{vanmegen1, pham,
sciortinotartaglia}. Many issues, however, still remain to be
understood, and this has prompted the proposal of extended versions
of this theory \cite{szamelmc, cao, reichman}, or the development of
alternative approaches \cite{tokuyama}.

In this context, in recent work a new first-principles theory of
dynamic arrest has been proposed \cite{rmf, todos1, todos2} which is
essentially the application of the self-consistent generalized
Langevin equation (SCGLE) theory of colloid dynamics
\cite{3,4,5,TAVSDB} to the analysis of dynamic arrest phenomena in
colloidal systems. The SCGLE theory was originally devised to
describe the tracer and collective diffusion properties of colloidal
dispersions in the short- and intermediate-times regimes. Its
self-consistent character, however, introduces a non-linear dynamic
feedback, leading to the prediction of dynamic arrest in these
systems, similar to that exhibited by the mode coupling (MC) theory
of the ideal glass transition. The resulting theory of dynamic
arrest in colloidal dispersions was applied in recent work to
describe the glass transition in three mono-disperse experimental
systems with specific (hard-sphere, screened electrostatic, and
depletion) inter-particle effective forces \cite{rmf, todos1}, with
the same quantitative predictive power as conventional MCT, but with
a smaller degree of difficulty in its application \cite{todos2}. The
present paper introduces the multi-component extension of the SCGLE
theory of colloid dynamics \cite{marco1,marco2} as a new alternative
first-principles theory of dynamic arrest in colloidal mixtures.

Colloidal mixtures offer a richer variety of possible dynamic arrest
scenarios \cite{stavans, imhofdhont1, imhofdhont2, vanmegen2,
moreno1, kikuchi}. These include, for example, the possibility that
the particles of one species become arrested, while the particles of
the other species continue to diffuse through the disordered matrix
of the arrested particles. In fact, an important prediction of the
multi-component extension of MCT \cite{bossethakur1, barratlatz1}
was precisely the existence of these mixed states \cite{thakurbosse,
bossekaneko, NAGELE2}. Later studies of dynamic arrest in mixtures
\cite{voigtmann1, voigtmann2, voigtmann3, voigtmann4, flennerzsamel,
voigtmann5} have been based on MCT not because it is the ultimate
and definitive theory, but because it was virtually the only
first-principles theory of dynamic arrest available that had a high
level of predictive capability. However, relevant issues remain that
have not been resolved by its application. For example, although the
one-component version of MCT predicts the existence of repulsive
glasses in a hard-sphere system, and its melting upon the addition
of a small amount of smaller particles (depletion forces), followed
by the reentrance to a new kind of (attractive) glass \cite{pham},
no reentrance seems to be predicted by the two-component version of
MCT \cite{zaccarellilowen}. Issues like this merit the development
of alternative approaches to the description of dynamic arrest
phenomena in mixtures, particularly if they are independent from
conventional MCT. Although the SCGLE theory presented here has
several features in common with MCT, such as its use of exact memory
function expressions for the concentration-concentration
time-correlation functions, in the SCGLE theory the derivations,
approximations and argumentation are conceived and based on a
totally different fundamental and conceptual framework
\cite{3,4,5,TAVSDB,marco1,marco2}.

Besides presenting the SCGLE theory of dynamic arrest in colloid
mixtures, in the present paper we illustrate the scenarios predicted
by this theory in the simplest conditions, namely, in its
application to model binary mixtures. For two such systems, a
repulsive Yukawa mixture and a hard-sphere mixture,  we numerically
solve the full self consistent system of dynamic equations for the
so-called propagators (or Green's functions) of the self and
collective intermediate scattering functions. We illustrate in this
manner the two possible dynamic arrest patterns possible in a binary
colloidal mixture, namely, the simultaneous and the sequential
arrest of the two species. In the latter case, we discuss the hybrid
or partially-arrested states in which one species is dynamically
arrested while the other species remains ergodic.

Determining the location of the ergodic--non-ergodic transitions and
calculating the non-ergodic parameters may be done by solving the
self-consistent system of equation or, more simply, by finding the
long-time asymptotic stationary (``fixed-point") solutions of the
self-consistent system of equations. Here we also address this
issue, and report the derivation of a strikingly simple and general
result which takes the form of a system of coupled equations for the
localization length of the particles of each species. Solving this
system of equations leads to the unambiguous identification of which
species is arrested (finite localization length) and which species
remains ergodic (infinite localization length). As a result, we are
able to draw the entire ergodic--non-ergodic phase diagram of the
binary hard-sphere mixture, which is one of the main specific
contributions of this work.

The present paper starts with a brief summary of the SCGLE theory of
the dynamics of colloidal mixtures. This theory was developed in
Ref.\ \cite{marco2} and tested there through its comparison with the
short- and intermediate-time results of Brownian dynamics computer
simulations for a model colloidal mixture interacting through
screened Coulomb pair interactions. That original version of the
SCGLE theory required as an input not only the partial static
structure factors, but also other equilibrium static structural
information associated with the exact short-time conditions that
this initial version of the theory aimed at incorporating. As it has
been shown more recently \cite{todos2}, however, a simplified
version of the SCGLE theory, in which this exact short-time
information is neglected, happens to be equally accurate,
particularly in the long-time regime and with respect to the
predicted dynamic arrest scenarios. Thus, we shall refer to this
simplified version simply as the SCGLE theory, which we review in
the following section.

Although our main interest here is in the description of dynamic
arrest phenomena, in section III we start our discussion by
reviewing the predictive capability of this simplified theory in the
short- and intermediate-time regimes of the model binary Yukawa
mixture studied in Ref.\ \cite{marco2} in one of the fully ergodic
states considered in that work. We then proceed to illustrate the
dynamic arrest scenarios predicted by this theory, by taking as the
initial reference the same repulsive Yukawa mixture, and increase
the intensity of the repulsion parameter (a process that amounts to
effectively lowering the temperature), until reaching a transition
to a dynamically arrested state. This turns out to be a transition
from a fully ergodic to a fully arrested state in which both species
are simultaneously arrested. In the second of our illustrative
examples, in section IV we consider a hard-sphere model colloidal
mixture, whose state parameters are the size disparity and the
concentration of the two species. Here we present an example of a
transition to a partially arrested state, in which the large
particles become arrested and the smaller particles continue to
diffuse in the disordered matrix formed by the arrested species. In
section V we derive the equations for the non-ergodic parameters of
the system, and illustrate its use in the context of the hard-sphere
binary mixture by determining its entire dynamic-arrest phase
diagram. The main conclusions of this paper are finally summarized
in section VI.

\section{The SCGLE theory of the dynamics of colloidal mixtures}

In this section we summarize the main elements of the self-
consistent theory of the dynamics of colloidal mixtures presented in
Refs.\ \cite {marco1, marco2}.  For a colloidal mixture with $\nu $
species, the dynamic properties can be described in terms of the
relaxation of the fluctuations $\delta n_{\alpha }({\bf r} ,t)$ of
the local concentration $n_{\alpha }({\bf r},t)$ of colloidal
particles of species $\alpha$ ($= 1, 2, ...\nu$) around its bulk
equilibrium value $n_{\alpha }$. The average decay of $\delta
n_{\alpha }({\bf r},t)$ is described by the time-dependent
correlation function $F_{\alpha \beta}(k,t)\equiv \left\langle
\delta n_{\alpha }({\bf k},t)\delta n_{\beta }(-{\bf
k},0)\right\rangle $ of the Fourier transform $\delta n_{\alpha
}({\bf k},t)\equiv (1/\sqrt {N_{\alpha }})\sum_{i=1}^N \exp {[i{\bf
k}\cdot{\bf r_i^{(\alpha) }}(t)]}$ of the fluctuations $\delta
n_{\alpha }({\bf r} ,t)$, with ${\bf r_i^{(\alpha) }}(t)$ being the
position of particle $i$ of species $\alpha $ at time $t$.
$F_{\alpha \beta}(k,t)$ is referred to as the partial intermediate
scattering function, measured by experimental techniques such as
dynamic light scattering \cite{1,2}. One can also define the $self$
component of $F_{\alpha \beta}(k,t)$, referred to as the
self-intermediate scattering function, as
$F_{\alpha\beta}^{(s)}(k,t)\equiv \delta_{\alpha \beta}\left\langle
\exp {[i{\bf k}\cdot \Delta{\bf R^{(\alpha )}}(t)]} \right\rangle $,
where $\Delta{\bf R}^{(\alpha )}(t)$ is the displacement of any of
the $N_{\alpha }$ particles of species ${\alpha}$ over a time $t$,
and $\delta_{\alpha \beta}$ is Kronecker's delta function.

For notational economy, let us denote the \emph{matrix} with
elements $F_{\alpha \beta}(k,t)$ ($\alpha, \beta = 1, 2, ..., \nu$)
by $F(k,t)$ or simply by $F(t)$, and similarly for $F^{(s)}(t)$.
These matrices have the initial values $F(t=0) = S$ and
$F^{(s)}(t=0) = I$, where $I$ is the identity matrix, $I_{\alpha
\beta} \equiv \delta_{\alpha \beta}$, and the elements of the matrix
$S$ are the partial static structure factors, $S_{\alpha
\beta}(k)\equiv \left\langle \delta n_{\alpha }({\bf k},0)\delta
n_{\beta }(-{\bf k},0)\right\rangle $. The matrix of intermediate
scattering functions may also be normalized by its initial value, to
define the \emph{collective propagator} matrix $\Psi(t) \equiv
F(t)S^{-1}$, which is such that $\Psi(0) = I$. In a similar manner,
we define the \emph{self propagator} matrix as $\Psi^{(s)}(t) \equiv
F^{(s)}(t)$, normalized to $\Psi^{(s)}(0) = I$.

The multi-component version of the SCGLE theory starts with the
exact time-evolution equation that governs the relaxation of the
Fourier transform (FT) $\delta n_{\alpha }({\bf k},t)$ of the
fluctuations of the local concentration of species $\alpha$. This
equation also governs the relaxation of the partial intermediate
scattering functions $F_{\alpha \beta}(k,t)$, and can be written as
an exact expression for the Laplace transform (LT) $F(z)$ of the
matrix $F(t)$, namely \cite{marco1},

\begin{equation}\label{fdz}
F(z)=\left\{z+(I+C(z))^{-1}k^{2}DS^{-1}\right\}^{-1}S,
\end{equation}
where the elements of the matrix $D$ are given by $D_{\alpha \beta}
\equiv \delta_{\alpha \beta} D^0_{\alpha}$, with $D^0_{\alpha}$
being the diffusion coefficient of species $\alpha$ in the absence
of interactions. This is related with the solvent friction
coefficient on an isolated particle of species $\alpha$,
$\zeta^0_{\alpha}$, through the Einstein relation,
$D^0_{\alpha}\equiv k_{B}T/\zeta^0_{\alpha}$. The elements
$C_{\alpha \beta}(k,z)$ of the matrix $C(z)$ are the LT of the
so-called irreducible memory functions $C_{\alpha \beta}(k,t)$
\cite{NAGELE2,noteirreducible}. The corresponding result for the
``self'' component, $F^{(s)}(z)$, of the matrix $F(z)$ reads

\begin{equation}\label{fsdz}
F^{(s)}(z)=\left\{z+(I+C^{(s)}(z))^{-1}k^{2}D\right\}^{-1},
\end{equation}
where the matrix $C^{(s)}(z)$ is the corresponding irreducible
memory function.

The second ingredient of the SCGLE theory is the intuitive notion
that collective and self dynamics may be connected in a simple
manner. Vineyard's approximation, in which $\Psi(t)$ is approximated
by $\Psi^{(s)}(t)$, is a rather primitive implementation of this
idea. The SCGLE theory incorporates the same notion, but implemented
at the level of the irreducible memory functions $C(t)$ and
$C^{(s)}(t)$, namely,

\begin{equation}
C(t)= C^{(s)}(t), \label{vineyard}
\end{equation}
which we shall simply refer as the Vineyard-like approximation. Let
us mention that in the original proposal of the multi-component
version of the SCGLE theory \cite{marco2} the difference
$[C(t)-C^{(s)}(t)]$ was approximated not by zero, as in this
equation, but by its exact (``single exponential") short-time limit
$[C^{SEXP}(t)-C^{(s),SEXP}(t)]$, given the fact that explicit
expressions exist \cite{sexp1, marco1} for the short-time limits
$C^{SEXP}(t)$ and $C^{(s),SEXP}(t)$ of the irreducible memory
functions $C(t)$ and $C^{(s)}(t)$ in terms of equilibrium static
structural properties of the system, which are considered known.
More recently, however, it was found \cite{todos2} that an equally
accurate but much simpler approximation results if the difference
$[C(t)-C^{(s)}(t)]$ is approximated by zero, as in Eq.
(\ref{vineyard}), thus eliminating the need to determine the
short-time memory functions $C^{SEXP}(t)$ and $C^{(s),SEXP}(t)$.

The third ingredient of the SCGLE theory was suggested \cite{marco2}
as the interpolation of $C^{(s)}(t)$ between its two exact limits at
small and large wave-vectors, by means of a phenomenological but
universal interpolation device. We know that the $C^{(s),SEXP}(t)$
is also the exact large-$k$ limit of $C^{(s)}(t)$, whereas the
small-$k$ limit is the so-called time-dependent friction function
matrix $\Delta\zeta^*(t)$. Thus, we write $C^{(s)}(k,t)=
C^{(s),SEXP}(k,t)+ [\Delta\zeta^*(t)-C^{(s),SEXP}(k,t)]\lambda(k)$,
where we have written explicitly the $k$-dependence of the various
matrices. The matrix $\lambda$ is a matrix of phenomenological
interpolating functions, whose elements are defined as
\cite{3,marco2}

\begin{equation}\label{lambda}
\lambda_{\alpha\beta}(k)\equiv
\delta_{\alpha\beta}/[1+(k/k^{(\alpha)}_{min})^2],
\end{equation}
with $k^{(\alpha)}_{min}$ being the position of the first minimum
following immediately the main peak of the partial static structure
factor $S_{\alpha\alpha}(k)$. At times long enough that the
short-time memory function $C^{(s),SEXP}(k,t)$ has already decayed,
the interpolation formula above for $C^{s}(k,t)$ simplifies still
further, reading

\begin{equation}
C^{(s)}(k,t)= [\Delta\zeta^*(t)]\lambda(k). \label{interpolation}
\end{equation}
As suggested in Ref.\ \cite{todos2}, it is this approximation that
enters in the present formulation of the multi-component SCGLE
theory of dynamic arrest.

The matrix $\Delta\zeta^*(t)$ in eq. (\ref{interpolation}) is a
diagonal matrix whose $\alpha$th diagonal element,
$\Delta\zeta_\alpha^*(t)$, is the time-dependent friction function
of particles of species $\alpha$. The final ingredient of the SCGLE
theory is an approximate but general result for this property,
which, as explained below, was derived as an application of the GLE
formalism \cite{faraday, delriocorrea, martinzon} to tracer
diffusion phenomena. Such an expression reads \cite{martinzon}

\begin{equation}
\Delta \zeta ^{*} _{\alpha}(t) =\frac{D^0_{\alpha}}{3(2\pi)^3}\int
d^3k k^2 [F^{(s)}(t)]_{\alpha\alpha} [c \sqrt{n} F(t) S^{-1}
\sqrt{n} h]_{\alpha\alpha}, \label{5}
\end{equation}
with the elements of the $k$-dependent matrices $h$ and $c$ being
the Fourier transforms $h_{\alpha\beta}(k)$ and $c_{\alpha\beta}(k)$
of the Ornstein-Zernike total and direct correlation functions,
respectively. Thus, $h$ and $c$ are related to $S$ by $S =
I+\sqrt{n}h\sqrt{n} = [I-\sqrt{n}c\sqrt{n}]^{-1}$, with  the matrix
$\sqrt{n}$ defined as $[\sqrt{n}]_{\alpha\beta} \equiv
\delta_{\alpha\beta}\sqrt{n_\alpha}$.

Let us now comment on the origin of Eq. (\ref{5}). In Ref.\ 
\cite{faraday} the effective Langevin equation of a tracer colloidal
particle interacting with the other particles of a mono-disperse
suspension was derived, using the concept of contraction of the
description \cite{delriocorrea} (a summary of such derivation is
contained in Appendix B of Ref.\ \cite{todos1}). Besides the solvent
friction force, $-\zeta^0 \textbf{V}(t)$, the direct interactions of
the tracer particle with the other particles give rise to an
additional friction term of the form $-\int_0^t dt'
\Delta\zeta(t-t')\textbf{V}(t') $, where $\textbf{V}(t)$ is the
tracer particle's velocity at time $t$. In the process, an exact
result for the time-dependent friction function
$\Delta\zeta^*(t)\equiv \Delta\zeta(t)/\zeta^0$ is generated which,
when complemented with two well defined approximations, leads to the
mono-disperse version of Eq. (\ref{5}). The multi-component
extension was derived in Ref.\ \cite{martinzon} from the exact result
in eq. (124) of that reference, upon the introduction of the
``homogeneous fluid" and the ``decoupling" approximations (eqs.
(135)-(138) and eq. (144) of the same reference).

Finally, let us mention that eqs. (\ref{fdz}) and (\ref{fsdz}) are
exact results, and that eq. (\ref{5}) derives from the exact result
just referred to. Hence, it should not be a surprise that the same
results are employed by other theories; in fact, these equations
also appear in the formulation of MCT. The difference lies, of
course, in the way we relate them and use them; in this sense, the
distinctive elements of our theory are then the Vineyard-like
approximation in eq. (\ref{vineyard}) and the factorization
approximation in eq. (\ref{interpolation}).

\section{Illustrative application: the model Yukawa mixture}

The set of coupled equations (\ref{fdz})-(\ref{5}) constitutes the
self-consistent generalized Langevin equation theory of the dynamics
of colloidal mixtures, and has to be solved numerically. For this,
equations (\ref{fdz}) and (\ref{fsdz}) are first Laplace-inverted,
and written as a set of coupled integro-differential equations
involving functions of $k$ and $t$. These functions are then
discretized in a mesh of points large enough to ensure independence
of the solution with respect to the size of the mesh. The
discretized system of equations is then solved by a straightforward
direct iteration method or, for the long times required in the study
of dynamic arrest phenomena, by the methods described in Refs.\ 
\cite{fuchsgoetze, banchiotesis}.

To illustrate its quantitative accuracy, the SCGLE theory was
applied in Ref.\ \cite{marco2} to the calculation of the dynamic
properties of a particular model system, namely, a binary mixture of
Brownian particles interacting through a hard-core pair potential of
diameter $\sigma$, assumed the same for both species, plus a
repulsive Yukawa tail of the form (in units of the thermal energy $\
k_{B}T=\beta ^{-1}$)

\begin{equation}
\beta u_{\alpha
\beta}(r)=A_{\alpha}A_{\beta}\frac{e^{-z(\frac{r}{\sigma}-1)}}{r/\sigma}.
\end{equation}

\noindent The dimensionless parameters that define the thermodynamic
state of this system are the total volume fraction $\phi \equiv
\frac{\pi}{6} n \sigma^3$ (with $n$ being the total number
concentration, $n=n_1+n_2$), the relative concentrations $x_{\alpha}
\equiv \frac{n_{\alpha}}{n}$ $(\alpha=1,2)$, and the potential
parameters $A_1$, $A_2$, and $z$. As in Ref.\ \cite{marco2}, here we
also assume that the free-diffusion coefficient $D^0_\alpha$ of both
species are identical, i.e., $D^0_1=D^0_2=D^0$. Explicit values of
the parameters $\sigma$ and $D^0$ are not needed, since the
dimensionless dynamic properties, such as $F_{\alpha \beta}(k,t)$,
only depend on the dimensionless parameters specified above, when
expressed in terms of the scaled variables $k\sigma$, and $t/t_0$,
where $t_0 \equiv {\sigma}^2/D^0$. In Ref.\ \cite{marco2} Brownian
dynamics simulations were reported for the static and dynamic
properties of this system with fixed screening parameter $z=0.15$
and various coupling parameters $A_1$ and $A_2$ and volume
fractions, in the short- and intermediate-time regimes.

\begin{figure}[ht]
\includegraphics[scale=.5]{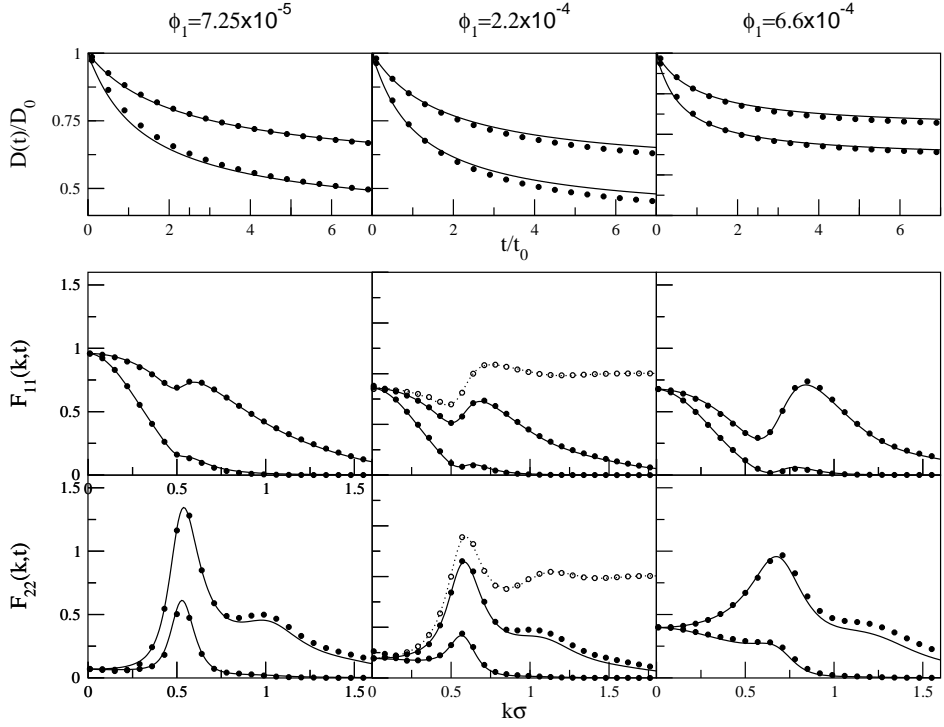}
\caption{ Time-dependent diffusion coefficients $D_\alpha(t)$ as a
function of time (upper curves corresponding to $D_1(t)$), and total
intermediate scattering functions $F_{11}(k,t)$ and $F_{22}(k,t)$ as
a function of wave-vector of a repulsive Yukawa mixture with
$z=0.15$, $A_1=10$, $A_2=10\sqrt{5}$ for $t=t_0$ and $t=10t_0$. The
volume fraction of the more interacting species is kept fixed at
$\phi_2=2.2\times 10^{-4}$ and $\phi_1$ takes the values
$\phi_1=0.725\times 10^{-4}$ (left column), $\phi_1=2.2\times
10^{-4}$ (middle column), and $\phi_1=6.6\times 10^{-4}$ (right
column). Solid lines are the SCGLE theoretical results and circles
are Brownian dynamics data. In the middle column we also include the
partial static structure factors $S_{\alpha \alpha}(k)\equiv
F_{\alpha \alpha}(k,t=0)$, with the dotted lines being a smooth
HNC-fit of the simulation data. \label{fig.1}}
\end{figure}

The partial static structure factors $S_{\alpha  \beta}(k)$ obtained
from those simulations are also employed here as the static input
needed in the implementation of the SCGLE theory. As a result, we
can calculate all the dynamic properties entering in the
self-consistent equations above, and other properties that derive
from them, such as the mean squared displacements $W_{\alpha
}(t)\equiv <[\Delta {\bf R}^{(\alpha )}(t)]^2>/6$ or the
time-dependent diffusion coefficients $D_{\alpha }(t)\equiv
W_{\alpha }(t)/t$. In Fig.\ \ref{fig.1} a comparison of these results
with the Brownian dynamics calculations is presented, involving the
case of a binary mixture with $z=0.15$, $A_1=10$, and
$A_2=10\sqrt{5}$. The volume fraction of the more interacting
species is kept fixed at $\phi_2=2.2\times 10^{-4}$ and $\phi_1$
takes the values $\phi_1=0.725\times 10^{-4}$ (left column),
$\phi_1=2.2\times 10^{-4}$ (middle column), and $\phi_1=6.6\times
10^{-4}$ (right column), corresponding to molar fractions
$x_1=0.25$, $x_1=0.5$ and $x_1=0.75$, respectively. This figure
corresponds to the same conditions as Fig. 3 of Ref.\ \cite{marco2},
and illustrates the fact that the present \emph{simplified} SCGLE
theory also provides an excellent description of the relaxation of
concentration fluctuations in colloidal suspensions in the short-
and intermediate-time regimes illustrated in the figure. We notice
that the largest departures of the theoretical results from the
simulation data appear only at the short times illustrated by the
time $t=t_0$, but the agreement improves at longer times, as
illustrated by the time $t=10t_0$. The main message of this figure
is that the present self-consistent theory of colloid dynamics
provides an adequate description of the dynamic properties of a
simple idealized mixture also in the short- and intermediate-time
regimes, even though it does not satisfy the exact short-time
conditions built in the original proposal of this theory
\cite{marco2}. With this confidence, let us now explore the
long-time regime under conditions typical of the vicinity of a
transition to dynamically arrested states.

For this, let us take the conditions of the middle column of Fig. 1
as a reference, and let us keep the parameters $\phi_1, \ \phi_2$
and $z$ fixed ($\phi_1=\phi_2=2.2 \times 10^{-4}$ and $z=0.15$)
while we increase the coupling parameters $A_1$ and $A_2$ in an
identical proportion, to mimic a process of lowering the
temperature. Thus, let us write $A_\alpha=A^0_\alpha\chi$
($\alpha=1,2$), with $A^0_1=10$ and $A^0_2=10\sqrt{5}$, and increase
$\chi$ from its unit value in Fig. 1 until a transition of dynamic
arrest is encountered. In this case, we do not run computer
simulations to provide the input partial static structure factors
needed in the self-consistent scheme. Instead, we resort to an
approximate liquid state theory, namely, the solution of the
Ornstein-Zernike integral equation within the hyper-netted-chain
(HNC) approximation \cite{hansen}. In doing this we only loose
quantitative precision, but no qualitative accuracy. In fact, the
HNC approximation will yield the same scenario, but at effective
values of the coupling parameters $A_1$ and $A_2$ larger than the
actual ones. For example, the fitting curves for the partial static
structure factors in the middle column of Fig. 1 were obtained using
the HNC approximation with effective coupling parameters given by
$A_\alpha=A^0_\alpha\chi$ with $\chi = 1.225$.

In Fig.\ \ref{fig.2} we present the results of the SCGLE theory for
the diagonal elements $\Psi_{\alpha\alpha}(k,t)$ of the collective
propagator matrix $\Psi(t)$ as a function of time for fixed
wave-vector $k=k_{max}$, where $k_{max}$ is the position of the main
peak of the corresponding static structure factor
$S_{\alpha\alpha}(k)$. This figure illustrates the fact that as we
lower the temperature starting from the reference state
($\chi=1.225$) the system reaches eventually a transition to an
arrested state. This occurs for $A_1=42.57$, and $A_2=95.19$, i.e.,
for $\chi=4.257$, and this is a transition from a fully ergodic to a
fully arrested state, i.e., both species become simultaneously
arrested. This is evidenced by the fact that below this threshold
condition, exemplified in the figure by the case $\chi=4.255$, both
propagators decay to zero, whereas above this condition, exemplified
by the case $\chi=4.330$, they decay to a finite non-ergodic value.
Thus, the non-ergodic parameters jump discontinuously at the
transition, from zero in the ergodic side, to finite values in the
glass side. Exactly the same behavior is exhibited by the
self-diffusion propagators $\Psi^{(s)}_{11}(k,t)$ and
$\Psi^{(s)}_{22}(k,t)$, not included in the present illustration.
Let us stress that this process of increasing $A_1$ and $A_2$
keeping $z$ fixed is a mathematical, rather than physical, exercise;
the practicable experimental manner to increase $A_1$ and $A_2$
would be to increase the particle's electric charge (or to lower the
dielectric constant of the solvent), which in reality would also
increase the screening parameter $z$. Here, however, we use this
exercise only to illustrate the pattern of dynamic arrest transition
in colloidal mixtures in which the particles of both species become
localized simultaneously.

Similar calculations can be carried out for other regions of the
parameter space ($\phi_1, \phi_2, A_1, A_2, z$), but this scanning
and its results in terms of  other possible regions that exhibit
interesting dynamic behavior constitutes an independent issue,
beyond our current objective of introducing our SCGLE theory of
dynamic arrest in mixtures and illustrating its use. We can mention,
however, that in other regions of this state space the theory
predicts also the other pattern of dynamic arrest, which we now
illustrate in the context of a simpler system, namely, the
hard-sphere colloidal mixture.

\section{The hard-sphere mixture}

Let us now consider the binary hard spheres model dispersion, i.e.,
the uncharged version of the previous system in which we now allow
for hard-sphere size disparity. Thus, we consider two species of
particles of hard core diameters $\sigma_1$ and $\sigma_2$, present
at concentrations $n_1$ and $n_2$. We use as dimensionless control
parameters the size asymmetry $\delta\equiv \sigma_1/\sigma_2 \leq
1$ and either the two volume fractions $\phi_1$ and $\phi_2$ (with
$\phi_\alpha \equiv \pi n_\alpha \sigma_\alpha^3/6$), or the total
packing fraction $\phi=\phi_1+\phi_2$ together with the molar
fraction of the smaller species, $x_1=n_1/(n_1+n_2)$. This system
serves to illustrate some features associated precisely with the
size disparity. As discussed in more detail below, one expects that
for mild asymmetries the only possible mode of dynamic arrest for
all volume fractions and all molar fractions is that in which both
species are arrested simultaneously. It is only in the regime of
disparate sizes that one expects a richer scenario, including the
second mode of dynamic arrest, in which the large particles are
first arrested while the other species continues to diffuse,
followed by a second transition at a higher total volume fraction in
which the smaller particles are finally localized inside the pores
of the matrix formed by the previously arrested particles. Thus,
there must be a threshold asymmetry $\delta_c$ beyond which this
might occur. One also expects, however, that even in the regime of
considerable size disparities but for conditions corresponding to a
few large particles in a see of smaller ones, the only mode of
dynamic arrest continues to be the mode of simultaneous dynamic
arrest of the two species, since in this case the arrest of the
small particles implies the arrest of the few large ones. We call
this the \emph{``chancaquilla"} limit \cite{chancaquilla}, in
reference to one of the many names of a rather universal sweet
prepared by the vitrification of liquid molasses in which nuts or
raisins are previously dispersed. Clearly, the dynamic arrest of the
supporting molasses implies the arrest of the nuts dispersed in it.
Hence, for large size asymmetries we expect to find both modes of
dynamic arrest located in the neighborhood of the opposite
composition limits, $x_1=0$ and $x_1=1$.

\begin{figure}[ht]
\includegraphics[scale=.3]{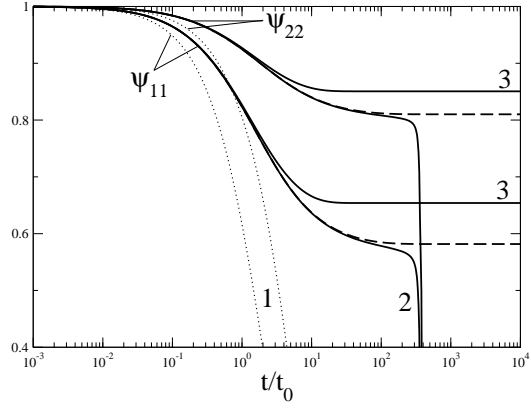}
\caption{SCGLE results for the collective diffusion propagators
$\Psi_{\alpha\alpha}(k_{max},t)$ at fixed wave-vector $k_{max}$ (the
position of the main peak of $S_{\alpha\alpha}(k)$) for the Yukawa
mixture with $z=0.15$, $\phi_1=\phi_2=2.2\times 10^{-4}$ calculated
using HNC partial static structure factors evaluated at effective
coupling parameters given by $A_\alpha=A_\alpha^0\chi$, with $
A_1^0=10$ and $A_2^0=10\sqrt{5}$, for values of $\chi$ corresponding
to the reference state in Fig. 1 ($\chi^{(1)}=1.225$, curves labeled
``1") and to states slightly below ($\chi^{(2)}=4.255$, labeled
``2") and above ($\chi^{(3)}=4.330$, labeled ``3"), the dynamic
arrest transition. The dashed plateaus indicate the value of the
non-ergodic parameters right at the transition ($\chi=4.257$).
\label{fig.2}}
\end{figure}

\begin{figure}[ht]
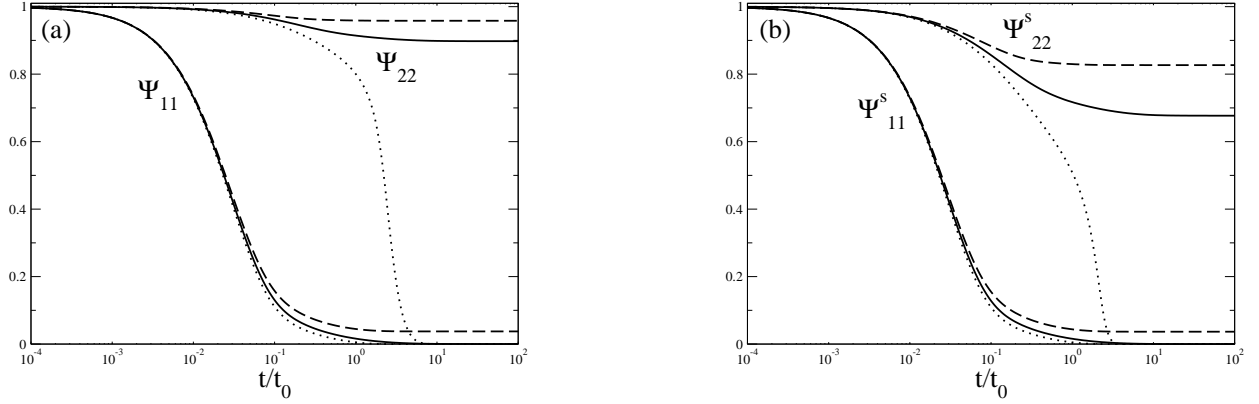

\begin{center}
\includegraphics[scale=.3]{fig3a.eps}\hfill
\includegraphics[scale=.3]{fig3b.eps}
\caption{(a) Collective diffusion propagators
$\Psi_{\alpha\alpha}(k_{max},t)$ and (b) self diffusion propagators
$\Psi_{\alpha\alpha}^{(s)}(k_{max},t)$ at fixed wave-vector
$k_{max}$ (the position of the main peak of $S_{\alpha\alpha}(k)$)
for the hard-sphere mixture with $\delta=0.3$ and $x_1=0.2$ for
different total packing fraction $\phi$. The results for $\phi=0.53$
(dotted lines) illustrate the fully ergodic states, whereas those
for $\phi=57$ (dashed lines) illustrate the fully arrested states,
in which both species are arrested. The results for $\phi=0.55$
(solid lines) illustrate the hybrid states in which the large
particles are arrested while the smaller particles continue to
diffuse.} \label{fig.3}
\end{center}
\end{figure}

In fact, this is exactly the scenario predicted by our theory. We
illustrate this with the SCGLE results obtained using the
Percus-Yevick \cite{lebowitz} structure factors as the static input,
for a size asymmetry $\delta=0.3 $. Let us calculate the collective
and self propagator matrices $\Psi(t)$ and $\Psi^{(s)}(t)$ at
increasing total volume fractions $\phi$ starting in the ergodic
regime, crossing the dynamic arrest transition line, and ending
inside the regime of arrested states. If this is done in the
neighborhood of the $x_1=1$ axis, say for $x_1=0.8$, the scenario
would be essentially that described by the \emph{chancaquilla}
mechanism involving the simultaneous dynamic arrest of both species,
with similar results to those discussed in Fig.\ \ref{fig.2} for the
Yukawa mixture. The other mode of dynamic arrest is observed in the
opposite regime, where the small particles are a minority,
exemplified by the molar fraction $x_1=0.2$. This case is
illustrated in Fig.\ \ref{fig.3}, which presents the evolution of the
relaxation of the diagonal elements of the collective and self
propagator matrices $\Psi(t)$ and $\Psi^{(s)}(t)$ as we increase the
total volume fraction $\phi$ at fixed molar fraction $x_1=0.2$,
starting in the ergodic regime. We find that, rather than passing
directly from the fully ergodic region to the region of fully
arrested states, as in Fig.\ \ref{fig.2}, one first passes through an
intermediate region of hybrid states, in which only the big
particles are arrested but the small particles continue to diffuse.
Thus, at a volume fraction $\phi_g^{(1)} \approx 0.545$, the system
passes from the fully ergodic region to the region $\phi_g^{(1)}  <
\phi <\phi_g^{(2)}$ of hybrid states, with $\phi_g^{(2)}$ being the
location of the second transition consisting of the localization of
the small particles in the pores of the random matrix of big
particles. In the present example this occurs at
$\phi_g^{(2)}\approx 0.555$.

Let us mention that in all these cases one observes that the
diagonal collective diffusion propagator $\Psi_{\alpha\alpha}(k,t)$
behave in qualitatively identical manner to the corresponding self
diffusion propagator $\Psi_{\alpha\alpha}^{(s)}(k,t)$ in the sense
that either both relax to zero or both exhibit dynamic arrest (i.e.,
decay to a finite value). One does not encounter a situation in
which, for example,$\Psi_{11}(k,t)$ decays to a finite value while
$\Psi_{11}^{(s)}(k,t)$ decays to zero. This indicates that at the
level of the diagonal propagators, self and collective diffusion
provide consistent descriptions of the pattern of relaxation. Thus,
we may identify the fully ergodic states by the fact that all the
elements of the (self and collective) propagators decay to zero,
whereas the fully non-ergodic states can be identified by the decay
of \emph{all} the propagators to finite asymptotic values
(identified with the non-ergodic parameters $\psi$ and $\psi^{s}$).
The results for $\phi=0.53$ in Fig.\ \ref{fig.3} are representative
of fully ergodic states, whereas those for $\phi=0.57$ illustrate
fully arrested states. Hybrid states are characterized by a mixed
behavior, in which the diagonal propagators $\Psi_{11}(k,t)$ and
$\Psi^{(s)}_{11}(k,t)$ associated to the mobile species, decay to
zero, while the diagonal propagators $\Psi_{22}(k,t)$ and
$\Psi^{(s)}_{22}(k,t)$, associated to the large particles, exhibit
arrested behavior. These conditions are illustrated in Fig.\ 
\ref{fig.3} by the results corresponding to $\phi=0.55$.

\begin{figure}[ht]
\begin{center}
\includegraphics[scale=.3]{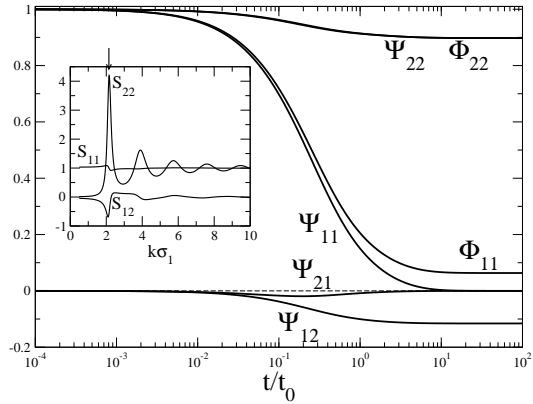}
\caption{Collective diffusion propagators $\Psi_{\alpha\beta}(k,t)$
and diagonal collective diffusion correlators
$\Phi_{\alpha\alpha}(k,t)$ at a common wave-vector corresponding to
the position of the main peak of $S_{22}(k)$ (see the inset, which
contains static structure factors at this state) for the hard-sphere
mixture with $\delta=0.3$ and $x_1=0.2$ for the hybrid state with
$\phi=0.55$.} \label{fig.3p}
\end{center}
\end{figure}

The relaxation of the crossed collective propagators
$\Psi_{12}(k,t)$ and $\Psi_{21}(k,t)$, not shown in Fig.\ 
\ref{fig.3}, do deserve some discussion, particularly in connection
with mixed states. First, one must have in mind that, although the
matrices $F(k,t)$ and $S(k)$ are symmetric, their product $F(k,t)
S^{-1}(k)=\Psi(k,t)$ in general is not; thus, in general,
$\Psi_{12}(k,t) \ne \Psi_{21}(k,t)$. Second, since the initial value
$\Psi(k,0)$ of the propagator matrix is the unit matrix, the initial
value of the off-diagonal collective propagators $\Psi_{12}(k,t)$
and $\Psi_{21}(k,t)$ must vanish. Third, the off-diagonal collective
propagators are not necessarily positive. This is illustrated in Fig.\ 
\ref{fig.3p}, which exhibits the time-relaxation of the four
collective propagators $\Psi_{\alpha\beta}(k,t)$ for the same mixed
state included in Fig.\ \ref{fig.3}, but now evaluated at a common
wave-vector $k$. The wave-vector employed in this figure corresponds
to the position of the maximum of $S_{22}(k)$; for reference, the
inset of this figure contains the partial static structure factors
of this system. At this wave-vector we observe that both,
$\Psi_{12}(k,t)$ and $\Psi_{21}(k,t)$, are negative, and that
$\Psi_{12}(k,t)$ relaxes to a finite non-ergodic asymptotic value,
while $\Psi_{21}(k,t)$ is always much smaller and relaxes to zero.

Let us mention that one could also display the same information not
in terms of \emph{propagators} but in terms of the so-called
collective \emph{``correlators"} $\Phi_{\alpha\beta}(k,t)$, defined
as $\Phi_{\alpha\beta}(k,t)\equiv
F_{\alpha\beta}(k,t)/S_{\alpha\beta}(k)$. Thus defined, the diagonal
collective correlator $\Phi_{\alpha\alpha}(k,t)$ is a linear
combination of the corresponding diagonal collective propagator
$\Psi_{\alpha\alpha}(k,t)$ and of a crossed propagator; more
precisely, $\Phi_{11}(k,t) =\Psi_{11}(k,t) + [S_{21}(k)/S_{11}(k)]
\Psi_{12}(k,t)$ and $\Phi_{22}(k,t) =\Psi_{22}(k,t) +
[S_{12}(k)/S_{22}(k)] \Psi_{21}(k,t)$. In the latter case, since
$\Psi_{21}(k,t)$ is small and relaxes to zero, we find that the
collective correlator $\Phi_{22}(k,t)$ is virtually identical to the
collective propagator $\Psi_{22}(k,t)$, as shown in Fig.\ 
\ref{fig.3p}. In contrast, although the propagator $\Psi_{11}(k,t)$
does decay to zero, the diagonal collective correlator
$\Phi_{11}(k,t)$ exhibits dynamic arrest due to its linear
dependence on $\Psi_{12}(k,t)$ which, as illustrated in the figure,
relaxes to a non-zero value. This observation may be relevant when
describing the hybrid states above. For example, the description in
terms of correlators has been employed in recent work \cite{NAGELE2,
moreno1, voigtmann5}, where the situation illustrated in Fig.\ 
\ref{fig.3p} arises, leading to the notion that collective diffusion
in mixtures is intrinsically different from self-diffusion. Our
results indicate, however, that this apparent difference is just the
result of the normalization convention employed in the definition of
the correlators, and that if the same information were displayed in
terms of the elements of the propagator matrices, no mixing of
relaxation modes would occur that suggest this notion. We should
stress that the \emph{self} correlators, defined as
$\Phi^{(s)}_{\alpha\beta}(k,t)\equiv F^{(s)}_{\alpha\beta}(k,t)$,
are identical to the self propagators
$\Phi^{(s)}_{\alpha\beta}(k,t)$, and hence correctly describe the
fact that only the small particles diffuse and the large particles
are arrested.

Finally, let us mention that distinguishing the three dynamic states
above (fully ergodic, fully non-ergodic, and hybrid states) with
only the information of the type in Figs.\ \ref{fig.3} and
\ref{fig.3p} is not a simple matter. The reason is that the
non-ergodic parameters of the propagators associated with the
localization of the small spheres do not jump discontinuously from
zero to a finite appreciable value at $\phi_g^{(2)}$. Instead, they
increase continuously from a value of zero right at this transition,
and hence, attain finite but very small values in the close
neighborhood of the transition. In the following section, however,
we shall derive a simpler procedure to precisely locate these
transitions and to evaluate the non-ergodic parameters in the
(partially or totally) arrested states.

\section{Fixed-point equations}

With the information just presented, one may choose to analyze the
detailed time-dependence of the propagators, and discuss the various
regimes of the relaxation processes described by these functions at
specific points in the state space. Alternatively, one can discuss
more global properties, such as the general topology of the
ergodic-nonergodic phase diagram of the system, just like one does
in equilibrium statistical thermodynamics. In fact, the
ergodic-nonergodic phase diagram should be an important complement
to conventional equilibrium phase diagrams in terms of practical
applications. If one is interested specifically in such global
properties, in principle one does not need to solve the set of
coupled equations (\ref{fdz})-(\ref{5}) for the full time- and
wave-vector-dependence of all the dynamic properties involved.
Instead, one may solve only the so-called fixed-point equations,
i.e., the equations for the long-time asymptotic stationary
solutions of the full SCGLE theory. The solutions thus determined
are precisely the non-ergodic parameters, whose zero or non-zero
value reveal if the state of the system is ergodic or dynamically
arrested.

In order to derive the fixed point equations, let us define the
non-ergodic parameter of the dynamic properties $\Psi(t)$,
$\Psi^{s}(t)$, $C(t)$, $C^{s}(t)$, and $\Delta \zeta ^* (t)$, as the
matrices $\psi (k)\equiv \lim_{t \to \infty}\Psi (k,t)$,
$\psi^{s}(k)\equiv \lim_{t \to \infty}\Psi^{s} (k,t)$, $c(k) \equiv
\lim_{t \to \infty}C (k,t)$, $c^{s}(k) \equiv \lim_{t \to
\infty}C^{s} (k,t)$, and $\Delta \zeta^{*(\infty)} \equiv \lim_{t
\to \infty}\Delta \zeta ^* (t)$, respectively. All the elements of
these matrices are zero in a fully ergodic state, and the non-zero
value of some or all of them indicate partial or total
non-ergodicity. Let us now substitute the dynamic properties
involved in the self-consistent system of equations in Eqs.\ 
(\ref{fdz})-(\ref{5}) by the sum of their asymptotic long-time value
above, plus the rest (which, by definition, always relaxes to zero).
In the resulting system of equations, let us then take the
asymptotic long-time limit, thus generating the following
self-consistent system of equations for the non-ergodic parameters

\begin{equation}
 \psi(k)=  [c(k) + k^2DS^{-1}(k)]^{-1}c(k), \label{nep1}
\end{equation}

\begin{equation}
 \psi^{s}(k)= [c^{s}(k) + k^2D]^{-1}c^{s}(k), \label{nep2}
\end{equation}

\begin{equation}
 c(k)= c^{s}(k),  \label{nep3}
\end{equation}

\begin{equation}
 c^{s}(k)= \lambda(k)\Delta \zeta^{*(\infty)}, \label{nep4}
\end{equation}

\noindent and

\begin{equation}
\Delta \zeta ^{*(\infty)} _{\alpha}
=\frac{D^0_{\alpha}}{3(2\pi)^3}\int d^3k k^2
[\psi^{(s)}]_{\alpha\alpha} [c \sqrt{n} \psi \sqrt{n}
h]_{\alpha\alpha}, \label{nep5}
\end{equation}

\noindent where the matrices $c$ and $h$ inside the integral in the
last equation refer to the direct and total Ornstein-Zernike
correlation functions, and together with the matrix $\sqrt{n}$, were
defined below Eq.\ (\ref{5}). Using Eqs.\ (\ref{nep3}) and
(\ref{nep4}) in Eqs.\ (\ref{nep1}) and (\ref{nep2}), we can express
the non-ergodic parameters $\psi(k)$ and $\psi_S(k)$ in terms of
$\Delta \zeta^{*(\infty)}$. Substituting the resulting expressions
in Eq.\ (\ref{nep5}), we finally get the following closed equation
for $\Delta \zeta^{*(\infty)}$

\begin{eqnarray}
\begin{split}
\Delta \zeta ^{*(\infty)} _{\alpha} = &
\frac{D^0_{\alpha}}{3(2\pi)^3}\int d^3k k^2 \left[ \left(\lambda(k)
+ k^2(\Delta
\zeta^{*(\infty)})^{-1}D\right)^{-1}\lambda(k)\right]_{\alpha\alpha}
\times \\
& \left\{c \sqrt{n} \left[ \left(\lambda(k) + k^2(\Delta
\zeta^{*(\infty)})^{-1}DS^{-1}(k)\right)^{-1}\lambda(k)\right]
\sqrt{n} h\right\}_{\alpha\alpha} ,
\end{split}
\label{nep5p}
\end{eqnarray}
for $\alpha=1,2,...,\nu$, with $\nu$ being the number of species.
Clearly, this equation always admits the trivial solutions
$\Delta\zeta^{*(\infty)}_\alpha=0$, which corresponds to the fully
ergodic fluid state. The existence of other non-zero real
solution(s) is associated to the existence of arrested states.

It is convenient to re-write Eq.\ (\ref{nep5p}) in terms of the
diagonal matrix $\gamma \equiv (\Delta \zeta^{*(\infty)})^{-1}D$,
whose diagonal elements $\gamma_\alpha$ will then satisfy the
following set of equations

\begin{equation}
\frac{1}{\gamma_\alpha}=\frac{1}{3(2\pi)^3}\int d^3k k^2 \left\{
\left[I + k^2\gamma\lambda^{-1}(k)
\right]^{-1}\right\}_{\alpha\alpha} \left\{c \sqrt{n} \left[ I +
k^2\gamma \lambda ^{-1}(k) S^{-1}(k)\right]^{-1} \sqrt{n}
h\right\}_{\alpha\alpha}. \label{nep5pp}
\end{equation}

\noindent For the same reasons given in the mono-component case
\cite{rmf, todos1}, the new order parameters $\gamma_\alpha$ may be
identified with the long-time asymptotic value of the mean squared
displacement (msd) of particles of species $\alpha$. Thus,
$\gamma_\alpha$ is infinite if that species diffuses, and is finite
if it is arrested; in fact, in the latter case $\gamma_\alpha^{1/2}$
is the localization length of the particles of that arrested
species. In terms of $\gamma$, Eqs.\ (\ref{nep1}) and (\ref{nep2})
for the non-ergodic parameters $\psi(k)$ and $\psi^{s}(k)$ read

\begin{equation}
\psi(k)= \left[ I + k^2\gamma \lambda ^{-1}(k)
S^{-1}(k)\right]^{-1}\label{nep1p}
\end{equation}

\noindent and

\begin{equation}
\psi^{s}(k)=  \left[I + k^2\gamma\lambda^{-1}(k) \right]^{-1}.
\label{nep2p}
\end{equation}

Eqs.\ (\ref{nep5pp})-(\ref{nep2p}) are the fixed point equations of
the SCGLE theory. In the particular case of a mono-disperse system,
one recovers the result derived and employed in Refs.\ \cite{rmf,
todos1}. For a given system, i.e., for given pair potentials, one is
supposed to determine first the static structure factors
$S_{\alpha\beta}(k)$, which is the only external input in Eq.\ 
(\ref{nep5pp}) (recall that the matrix $\lambda(k)$ is determined by
these structural properties, according to Eq.\ (\ref{lambda})). Eqs.\ 
(\ref{nep5pp}) can be solved numerically, and the solutions for
$\gamma$ are then employed in Eqs.\ (\ref{nep1p}) and (\ref{nep2p})
to determine the non-ergodic parameters.

Let us apply Eqs.\ (\ref{nep5pp}) to calculate the order parameters
$\gamma_1$ and $\gamma_2$ for binary hard-sphere mixtures. The
corresponding partial static structure factors $S_{\alpha \beta}(k)$
needed in that equation will again be provided by the Percus-Yevick
(PY) approximation. From the resulting values of $\gamma_1$ and
$\gamma_2$, we can classify the states of this system as ergodic,
fully non-ergodic, or partially non-ergodic, and thus scan the state
space to determine the regions of these dynamic phases and the
boundaries between them. For this, let us adopt the asymmetry
parameter $\delta$, the total volume fraction $\phi$, and the molar
fraction $x_1$ of the smaller spheres as the dimensionless state
variables spanning the state space $(\delta, \phi, x_1)$.

\begin{figure}[ht]
\begin{center}
\includegraphics[scale=.3]{fig5a.eps}\hfill
\includegraphics[scale=.3]{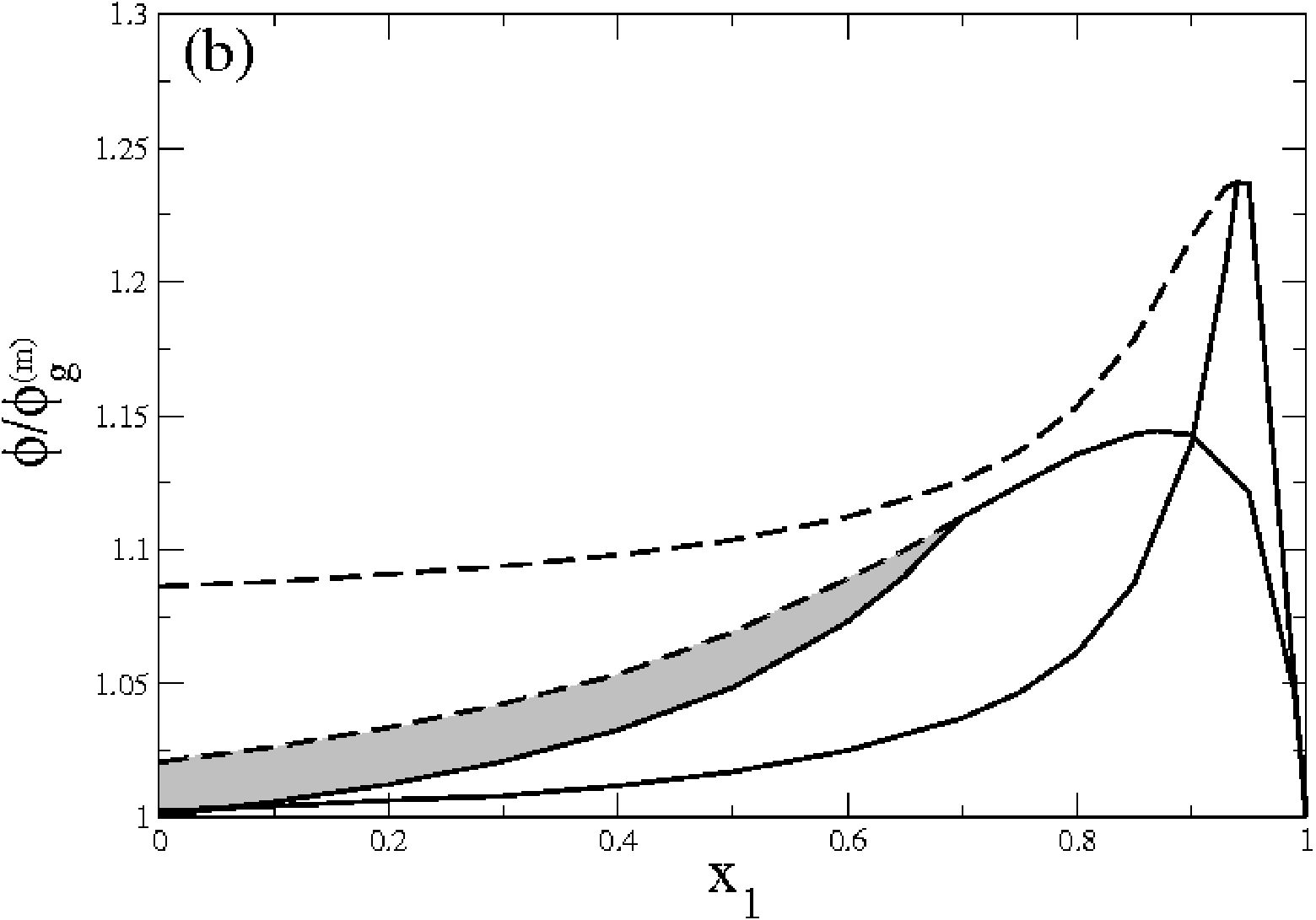}
\caption{Phase diagram of dynamic arrest states of the binary
hard-sphere mixture. The three-dimensional state space ($\delta,
\phi, x_1$) is spanned by the size disparity parameter $\delta\equiv
\sigma_1/\sigma_2\le 1$, the total volume fraction $\phi \equiv
\phi_1 + \phi_2$ with $\phi_\alpha \equiv \pi n_\alpha \sigma_\alpha
^3/6$, and the molar fraction of the smaller particles $x_1\equiv
n_1/(n_1+n_2)$, where $\sigma_\alpha$ and $n_\alpha$ are the
hard-core diameter and the number concentration of particles of
species $\alpha$. This state space is partitioned in three regions,
corresponding to fully ergodic, mixed, and fully non-ergodic states.
In this figure we show the intersection lines of this transition
surface with the planes of constant $\delta$. The total volume
fraction $\phi$ has been scaled with the volume fraction
$\phi_g^{(m)}$ of the ideal glass transition of the mono-disperse
hard-sphere system. In (a) we illustrate the regime of mild size
disparities with the values $\delta = 1.0$ (horizontal line,
corresponding to $\phi=\phi_g^{(m)}$), 0.8, 0.6 (thicker line), and
0.4, where the respective line divides the subspace ($\phi, x_1$) in
the regions of fully ergodic states (below the line) and fully
non-ergodic states. The four dots linked by the arrow at fixed total
volume fraction $\phi / \phi_g^{(m)}= 1.0175$ correspond to the
states prepared in the experiment of Williams and van Megen (Ref.\
\cite{vanmegen2}) for $\delta=0.6$. In (b) we illustrate the regime
of severe size-disparities with the values $\delta = 0.3 $ and 0.2.
In this regime the region of fully ergodic states also lies below
the solid line but a region of mixed states appear (shaded region
for the case $\delta = 0.3)$. This region is bounded from above by
the localization transition of the small particles (dashed lines).}
\label{fig.4}
\end{center}
\end{figure}

Before presenting the results of this exercise, that we collect in
Fig.\ \ref{fig.4}, let us say that from intuitive arguments one may
advance some of the features of this phase diagram. For example, as
already indicated in section IV one expects that for mild
asymmetries the only possible mode of dynamic arrest for all volume
fractions and all molar fractions is that in which both species are
arrested simultaneously. The reason for this expectation is that
this regime must contain the extreme case corresponding to the
degenerate limit $\delta = 1$. From the point of view of their
interactions, in this limit the particles of both species are
actually indistinguishable ($\sigma_1=\sigma_2$), thus being in fact
a mono-disperse system, whose particles have been classified only
artificially as belonging to either of these two species. We expect,
then, that for any molar fraction $0\le x_1 \le1$, the system will
be arrested at the total volume fraction $\phi_g^{(m)}$ at which a
mono-disperse system would be arrested. This means that in the plane
$(\delta=1, \phi, x_1)$, the transition line must correspond to the
condition $\phi_1 + \phi_2 = \phi_g= \phi_g^{(m)}$. According to the
mono-component version of the SCGLE theory, and within the same
level of approximation adopted here for the static structure factor
(i.e., the PY approximation) we know that $\phi_g^{(m)}=0.537$
\cite{todos1}. Thus, in the degenerate case the straight line $\phi=
\phi_g^{(m)}$ (horizontal line in Fig.\ \ref{fig.4}.a) separates the
ergodic region $\phi < \phi_g^{(m)}$ from the region $\phi >
\phi_g^{(m)}$ of totally non-ergodic states, in which both species
are dynamically arrested.

Thus, the first question is now if the SCGLE theory has built-in
this degenerate limit, i.e., if under these degenerate conditions,
Eqs.\ (\ref{nep5pp})-(\ref{nep2p}) reduce to the mono-disperse result
discussed in references \cite{rmf, todos1}. Indeed, if one
constrains the solution for $\gamma_1$ and $\gamma_2$ to exhibit
this symmetry, i.e., that $\gamma_1=\gamma_2$, one can show that the
two equations involved in Eq.\ (\ref{nep5pp}) are indeed identical to
each other and to the corresponding mono-disperse equation,
independently of the composition represented by the molar fraction
$x_1$. This means that both ``species" either remain ergodic or are
arrested together.

The next question is then what happens when this degeneration is
broken, so that $\sigma_1 \le \sigma_2$, but remaining in the region
of mild asymmetry $\delta \lesssim 1$, in which we expect small
deviations from the degenerate limit. Unfortunately, we do not have
good intuitive arguments to anticipate even the sign of these
deviations. In Fig.\ \ref{fig.4}.a, however, we report the
predictions of our theory for this regime, in terms of the glass
transition lines corresponding to the values of the asymmetry
parameter $\delta = 1.0,\ 0.8,\ 0.6$ and 0.4, for which our theory
predicts that the glass transition line moves to higher total volume
fractions $\phi_g$ as the size disparity increases (i.e., $\delta$
decreases), except at the two ends $x_1=0$ and $x_1=1$ of the
transition line, where the value of $\phi_g$ must be fixed by its
mono-disperse value $\phi_g^{(m)}$ for all size disparities.

The general trends illustrated by Fig.\ \ref{fig.4}.a were actually
first observed and described by Williams and van Megen
\cite{vanmegen2} in their experimental process of melting an
originally mono-disperse glass by the replacement of a fraction of
its particles by particles of a smaller size keeping the same total
volume fraction. The dots and the arrow in Fig.\ \ref{fig.4}.a
correspond to their experimental process, that may conceptually be
described as driving a system from the initially mono-disperse glass
state $(\delta,\ \phi/\phi_g^{(m)},\ x_1)=(0.6,\ 1.0175,\ 0.0 )$
through the arrested states $(0.6,\ 1.0175,\ 0.196)$ and $(0.6,\
1.0175,\ 0.340)$, to end in the ergodic state $(0.6,\ 1.0175,\
0.537)$ which lies in the ergodic region according to our
predictions for the same size asymmetry $\delta=0.6$ of the
experimental system (the representation of this process in the scale
of Fig.\ \ref{fig.4}.a involves the re-scaling of the experimental
volume fractions $\phi^{exp}=0.58$ with the experimental value of
$\phi_g^{(m)}$ which, within the experimental errors, we took as
$\phi_g^{(m)}=0.57$).

The regime illustrated in Fig.\ \ref{fig.4}.a was also studied by
G\"otze and Voigtmann with MCT \cite{voigtmann1,voigtmann2}, with
predictions qualitatively similar to ours for intermediate
size-disparities ($ \delta \lesssim 0.65$), but with conflicting
predictions for milder size disparities ($ \delta \sim 0.8$), where
MCT predicts ``S-shaped" transition lines that may go below the
mono-disperse glass transition volume fraction $\phi_g^{(m)}$
\cite{voigtmann1}. The SCGLE theory, predicts, instead, that the
transition lines for the mixture, as illustrated in Fig.\ 
\ref{fig.4}.a always lie above the horizontal transition line
corresponding to the degenerate mono-disperse case. These
conflicting predictions of MCT and the SCGLE theory could be
resolved by experiments and/or computer simulations. The actual
magnitude of the difference between the predictions of the two
theories is, however, so small that a conclusive discrimination
between them will require experimental or simulation results of
higher resolution than provided by the currently available data
\cite{vanmegen1, voigtmann3}.

Fig.\ \ref{fig.4}.b illustrates the regime of large size disparities.
The SCGLE theory predicts that below a threshold asymmetry $\delta_c
\sim 0.4$, a region of mixed states appear, which is illustrated by
the shaded area in the results for $\delta=0.3$. This region is
bounded on the left by the axis $x_1=0$, below by the transition of
dynamic arrest of the large particles (solid line), and above by the
localization transition of the small particles (dashed line). The
region of mixed states ends on the right in the bifurcation point
where these two transition lines meet, and where the transition of
simultaneous dynamic arrest of both species starts, ending in the
axis $x_1=1$ (solid line). The area of the shaded region increases
as $\delta$ decreases, as illustrated by the results for $\delta =
0.2$, in which the region of mixed states was not shaded. Of course,
as $\delta$ approaches the threshold value $\delta_c \sim 0.4$, the
bifurcation point moves to the axis $x_1=0$, and the region of mixed
states disappears. Although we cannot in practice explore the limit
$\delta << 1$, we do not expect additional qualitatively different
trends beside those exhibited by the illustrative cases in Fig.\ 
\ref{fig.4}. Thus, to our knowledge, this is the first time that the
entire ergodic--non-ergodic phase diagram of a binary colloidal
mixture has been outlined.

\section{Concluding remarks}

In this paper we have introduced the self-consistent generalized
Langevin equation theory of dynamic arrest in colloidal mixtures,
and we have applied it to the description of dynamic arrest
phenomena in two simple model systems, namely, a repulsive Yukawa
and a hard-sphere mixtures, for which we illustrated the scenarios
predicted by this theory through the full numerical solution of the
coupled system of dynamic equations for the propagators $\Psi(t) $
and $\Psi^{(s)}(t)$ of the self and collective intermediate
scattering functions. In this manner we identified and illustrated
the two possible patterns of dynamic arrest in a binary colloidal
mixture, namely, the simultaneous and the sequential arrest of the
two species. The former corresponds to the transition from the
region of fully ergodic states, characterized by infinite values of
the order parameters $\gamma_1$ and $\gamma_2$, to the region of
fully non-ergodic states, characterized by finite values of these
two parameters. This transition is characterized by a discontinuous
jump of both order parameters. This is the only dynamic arrest
transition possible for moderate size asymmetry in a binary
hard-sphere mixture, as illustrated in Fig.\ \ref{fig.2}. It is also
the only transition possible for large asymmetries in the case of
hard spheres provided the system is in the limit of a few large
particles dispersed in a sea of small particles (the dynamic arrest
of ``chancaquilla" limit).

The second pattern of dynamic arrest is only observed for large
enough size disparities starting in the regime in which the small
particles are a minority. As illustrated in Fig.\ \ref{fig.3}, here
the system may be driven from the region of fully ergodic states
($\gamma_1=\gamma_2=\infty$) to the region of mixed states (finite
$\gamma_2$ and $\gamma_1=\infty$) in which the large particles are
arrested and the small ones diffuse through the voids left by the
large species. In this transition $\gamma_2$ jumps discontinuously
from its infinite value in the fully non-ergodic region to a finite
value in the region of mixed states, whereas $\gamma_1$ retains its
infinite value. Finally, there is a transition from this region of
mixed states to the region of fully ergodic states in which also the
small particles are localized (finite values of both, $\gamma_1$ and
$\gamma_2$). This latter transition is characterized by a continuous
change in the values of both order parameters $\gamma_1$ and
$\gamma_2$, although the latter changes continuously from its
infinite value inside the region of mixed states and right at the
transition, to finite values inside the fully non-ergodic region.

Locating the transitions just described may be done by numerically
solving the full self-consistent system of equation, as done in
Figs.\ \ref{fig.2} and \ref{fig.3}, or by directly determining the
non-ergodic parameters, which play the role of order parameters of
the transition. The latter can be done more economically through the
solution of the ``fixed-point" equations. In section IV we derived a
strikingly simple and general result for the non-ergodic parameters
$\gamma_1$ and $\gamma_2$, whose use was illustrated with the
construction of the phase diagram of the hard-sphere binary mixture.
We must say that some of the features of the SCGLE results thus
determined were in fact discovered and discussed in the pioneering
work of Refs.\ \cite{bossethakur1, thakurbosse, bossekaneko, NAGELE2}
based on the multi-component version of MCT. Our aim here has been
to document what is the scenario offered by the SCGLE theory for the
same conditions previously studied, and to extend the description to
new aspects not yet discussed with MCT or with other theoretical
approaches. In this regard, the main contribution of the present
work is the proposal of a new theoretical approach that allows the
determination of the entire dynamic phase diagram of simple model
colloidal mixtures. Thus, the second important contribution involves
the simplest of such models, namely, the hard-sphere binary mixture,
for which we outlined the entire dynamic arrest phase diagram. In
this context we discussed the regime of mild size asymmetries, in
which we reported the first apparent difference with the results of
MCT, namely, the fact that the SCGLE theory does not predict the
S-shaped transitions predicted by G\"otze and Voigtmann
\cite{voigtmann1} on the basis of MCT. For more severe size
disparities we also illustrated the main features of the SCGLE
predictions, including the precise location of the region of mixed
states, and its emergence for size disparities beyond a threshold
asymmetry $\delta_c \approx 0.4$. These results do not have yet a
MCT counterpart to compare with.

In this paper we also discussed the fact that other properties
besides $\gamma_1$ and $\gamma_2$ could be employed as order
parameters to classify the various dynamic phases in a mixture. For
example, one can use the diagonal elements of the non-ergodic
parameters associated with the self or the collective propagator
matrices, which also determine unambiguously the various possible
states. Although not central to the main subject of this paper, we
also explained that the use as order parameters of collective
\emph{correlators}, defined as the partial intermediate scattering
functions divided by their initial value, may lead to a confusing
description of the hybrid states in which one species is arrested
and the other remains ergodic.

In summary, we have presented a theory for dynamic arrest phenomena
in colloidal mixtures that offers some advantages over the extension
to mixtures of conventional MCT. First, it derives from a simpler
conceptual formalism, which might allow possible generalizations and
extensions. Second, the resulting equations are much simpler to
solve in practice than the corresponding equations of MCT. In fact,
the determination of (ergodic--non-ergodic) phase diagrams, such as
that discussed in the previous section, is considerably simplified
in the present case through the use of the results for the
non-ergodic parameters $\gamma_\alpha$ in Eq.\ (\ref{nep5pp}). Thus,
many important issues regarding the dynamics and the
(ergodic--non-ergodic) phase behavior of colloidal mixtures may be
discussed with the assistance of the SCGLE theory just presented.
The implications of the results presented here in the context of
specific physical phenomena and experimental conditions will,
however, be the subject of separate reports.

\section{ACKNOWLEDGMENTS:}
 This work was supported by the Consejo
Nacional de Ciencia y Tecnolog\'{\i}a (CONACYT, M\'{e}xico), through
grant No. SEP-2004-C01-47611 and of FAI-UASLP. The authors are
deeply indebted to Profs. J. Bergenholtz, A. Banchio, and G.
N\"agele for useful discussions.

\newpage
\section*{FIGURES AND CAPTIONS:}

\bigskip

\bigskip

\begin{thebibliography}{99}
\bibitem{chancaquilla} Chancaquilla is one of the many names for
one of the many versions of this candy, in this case made with
pumpkin and sesame seeds in the Rio Verde region of the mexican
state of San Luis Potos\'i.
\end{thebibliography}
\end{document}